\documentclass[a4paper]{jpconf}

\usepackage{graphicx}
\usepackage{amsmath}
\usepackage{amsfonts}
\usepackage{amssymb,latexsym}
\usepackage[english]{babel}
\usepackage{mathcomp}
\usepackage{dsfont}
\usepackage{array}
\newcommand{\dd}{\mathrm{d}}
\def\be{\begin{equation}}
\def\ee{\end{equation}}
\def\bea{\begin{eqnarray}}
\def\eea{\end{eqnarray}}
\def\fnl{f_\mathrm{NL}}
\def\dd{\mathrm{d}}
\def\lu{\ell_1}
\def\ld{\ell_2}
\def\lt{\ell_3}

\begin{document}

\title{Non-Gaussianity from extragalactic point-sources} 
\author{Fabien Lacasa}
\address{Institut d'Astrophysique Spatiale (IAS), B\^atiment 121, F-91405 Orsay
(France); Universit\'e Paris-Sud 11 and CNRS (UMR 8617)}
\ead{fabien.lacasa@ias.u-psud.fr}

\begin{abstract}
The population of compact extragalactic sources contribute to the non-Gaussianity (NG) at Cosmic Microwave Background (CMB) frequencies. We study their NG using publicly available full-sky simulations. We introduce a parametrisation to visualise efficiently the bispectrum and we describe the scale and frequency dependences of the bispectrum of radio and IR point-sources (PS). We show that the bispectrum is well fitted by an analytical prescription. We find that the clustering of IR sources enhances their NG by several orders of magnitude, and that their bispectrum peaks in the squeezed triangles. Examining the impact of these sources on primordial non-Gaussianity estimation, we find that radio sources yield an important positive bias to local $\fnl$ at low frequencies but this bias is efficiently reduced by masking detectable sources. IR sources produce a negative bias at high frequencies, which is not dimmed by the masking as their clustering is dominated by faint sources.
\end{abstract}

\vspace{-0.7cm}
\section{Introduction}
The CMB, the dominant signal on the sky around 100 GHz, is a powerful probe of the early universe which significantly contributes to the establishment of the standard model of cosmology. In the 30-350 GHz frequency range, other signals contribute at small angular scales and modify the statistical distribution (initially close to Gaussian) of the measured CMB anisotropies. Three of these signals are associated with extragalactic sources:\\
- galaxy clusters, in which the electrons of the hot ionised gas scatter off the CMB photons leaving a distinct spectral signature (Sunyaev-Zeldovich effect)\\
- radio loud galaxies with Active Galactic Nuclei emitting through synchrotron and free-free processes\\
- dusty star-forming galaxies, where the UV emission from stars heats the dust which consequently reemits in the infrared domain

We will focus on the characterisation of the distribution of these two latter populations. Radio sources can be considered randomly distributed on the sky and are hence modeled as a white-noise entirely described by the sources number counts. Their harmonic ``N-point" correlation functions (power spectrum, bispectrum, etc) are then flat and related to the corresponding moment of the 1-point distribution. By contrast, IR sources are highly clustered therefore their harmonic correlation functions require the inclusion of the spatial distribution of the sources in addition to their number counts.

The primary NG in the CMB can break the degeneracy between the primordial processes generating the cosmological perturbations e.g. the inflations models. While most models predict similar power spectra, they may be distinguished at the 3-point or higher level. As an example, standard single-field models predict undetectably small NG while multifield models, non-canonical kinetic terms or vacuum initial condition etc produce potentially detectable NG. As of today, no definite deviation from Gaussianity has been found in the CMB data, even if there is a hint of `local'-type NG in WMAP7 data \cite{Komatsu2011}. The local-type NG, produced generically by several inflation models, takes the form :
\be\label{Eq:fnldef}
 \Phi(\mathbf{x}) = \Phi_G(\mathbf{x}) + \fnl \left(\Phi_G(\mathbf{x})^2- \langle\Phi_G(\mathbf{x})^2 \rangle\right)
 \ee
where $\Phi$ is the Bardeen potential and the subscript G denotes the Gaussian linear part.

To study the distribution of radio and IR galaxies we use publicly available full-sky simulated maps of these sources \cite{Sehgal2010} between 30 and 350 GHz at high angular resolution. The results presented in this communication are presented in greater details in \cite{Lacasa2011}

\section{Bispectrum}
\begin{figure}[htbp]
\begin{center}
\includegraphics[width=7cm]{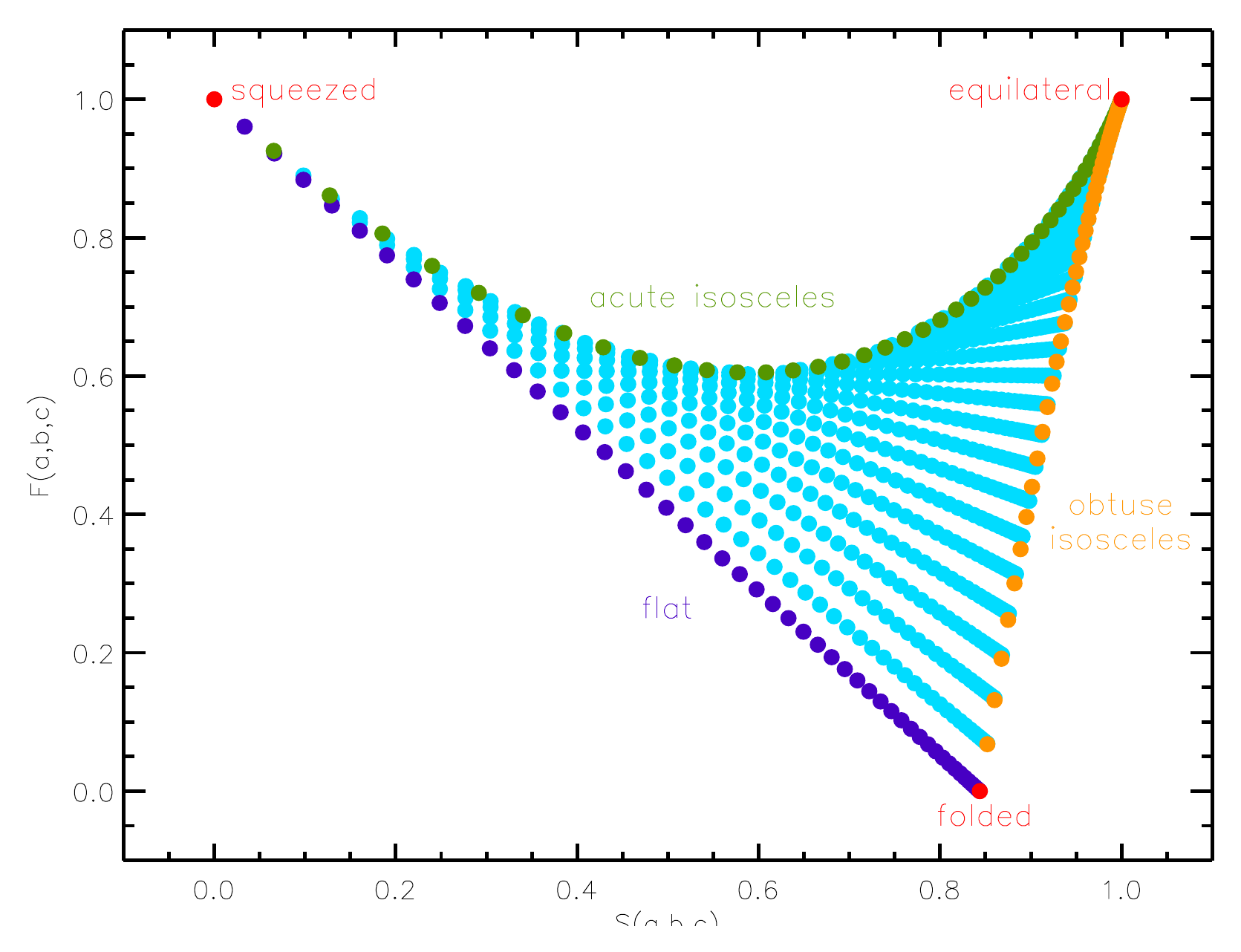}
\hspace{0.5cm}
\includegraphics[height=5.5cm]{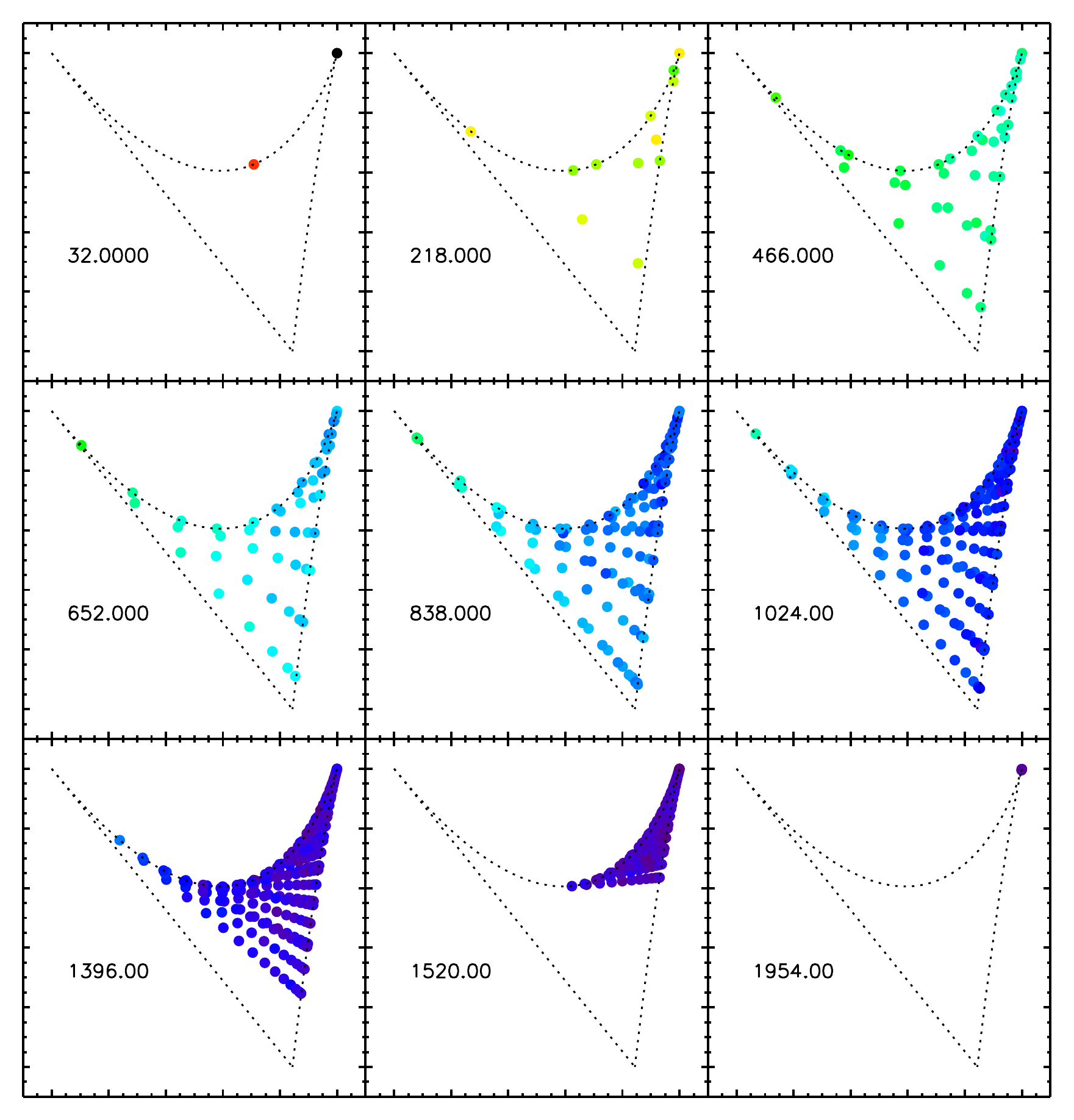}
\hspace{0.5cm}
\begin{minipage}[b]{3cm}
\caption{\label{Fig:triparametparamIR} \textbf{Left panel:} Positions of the triangle of a given perimeter in the (F,S) plan of the parametrisation.\\ \textbf{Right panel:} IR bispectrum at 148 GHz in our parametrisation}
\end{minipage}
\end{center}
\end{figure}

A gaussian field on the sphere is entirely characterised by its mean and 2-point correlation function (or its power spectrum). The 3-point correlation function is thus the lowest order, and most prominent, indicator of NG. The bispectrum is this 3-point correlation function in harmonic space :
\be
\langle a_{\ell_1 m_1} a_{\ell_2 m_2} a_{\ell_3 m_3}\rangle = G_{\ell_1 \ell_2 \ell_3}^{m_1 m_2 m_3} \times b_{\ell_1 \ell_2 \ell_3} \quad \mathrm{with} \quad G_{\ell_1 \ell_2 \ell_3}^{m_1 m_2 m_3} = \int \dd^2 n \, Y_{\ell_1 m_1}(n) Y_{\ell_2 m_2}(n) Y_{\ell_3 m_3}(n)
\ee
A full calculation of the bispectrum at WMAP or Planck resolution ($\ell_\mathrm{max} \sim 700$ and $\ell_\mathrm{max} \sim 2000$ respectively) is computationally very intensive so one usually resort to binning multipoles. \\
The bispectrum is invariant under permutation of $(\ell_1,\ell_2,\ell_3)$, it is a function of the triangle shape only. Three triangles shapes are of particular importance : squeezed (one side much smaller than the two others), equilateral and folded (flat isosceles).
To plot a bispectrum in a efficient way, e.g. without redundancy of information, we introduced a parametrisation accounting for this permutation invariance, and we defined three parameters (P,F,S) based on the elementary symmetric polynomials $\sigma_1=\lu+\ld+\lt$, $\sigma_2=\lu\ld+\lu\lt+\ld\lt$ and $\sigma_3=\lu\ld\lt$. The locations of the different triangles of constant perimeter P in the (F,S) plan is shown in Fig.\ref{Fig:triparametparamIR} (left panel). One can then plot a bispectrum by making slices of perimeters and colour-coding the value of the bispectrum.

\vspace{-0.3cm}

The local-type primordial NG defined in Eq.\ref{Eq:fnldef} leaves a characteristic imprint on the CMB, which can be detected at the 3-point level. The bispectrum thus generated peaks in the squeezed triangles, and because it has a separable form \cite{Komatsu2005} it has been shown that a fast estimator for its amplitude $\fnl$ can be built, which does not need the --prohibitive-- computation of the full bispectrum. This, commonly-called KSW, estimator is optimal in the sense that it minimizes the $\chi^2$ of the fit of the observed bispectrum to the local bispectrum :
\bea
\hat{f}_\mathrm{NL} = \sigma^2(\fnl) \sum_{\lu\leq\ld\leq\lt} N_{\lu\ld\lt} \frac{b_{\lu\ld\lt}^\mathrm{obs}\, b_{\lu\ld\lt}^\mathrm{loc}}{C_{\lu} C_{\ld} C_{\lt}} \label{Eq:KSW}\\
\mathrm{with} \quad \sigma^2(\fnl) = \sum_{\lu\leq\ld\leq\lt} N_{\lu\ld\lt} \frac{\left(b_{\lu\ld\lt}^\mathrm{loc}\right)^2}{C_{\lu} C_{\ld} C_{\lt}}
\eea
where $N_{\lu\ld\lt} = \frac{(2\ell_1+1)(2\ell_2+1)(2\ell_3+1)}{4\pi} \begin{pmatrix} \ell_1 & \ell_2 & \ell_3 \\ 0 & 0 & 0 \end{pmatrix}^2$ is the number of such configurations on the sky

\section{Point-sources}
Radio sources are randomly distributed on the sky, so they produce a constant bispectrum $b_\mathrm{ps} \propto \int S^3 \frac{\dd N}{\dd S} \dd S$, but as higher frequencies are surveyed e.g. with Planck, the IR population becomes of growing importance and one must account for the superposition of these two signals. From a method proposed in \cite{Argueso2003}, we have developped a full-sky analytical prescription to compute the bispectrum and higher order moments based on number counts and power spectrum. The prescription can handle individual populations and their superposition \cite{Lacasa2011}. Analytically, at the bispectrum level, it yields :
\be
b^\mathrm{IR}_{\lu\ld\lt} = \alpha \sqrt{C^\mathrm{IR}_{\lu} C^\mathrm{IR}_{\ld} C^\mathrm{IR}_{\lt}} \quad \mathrm{with} \quad \alpha = \frac{\int S^3 \frac{\dd N}{\dd S} \dd S}{\left(\int S^2 \frac{\dd N}{\dd S} \dd S\right)^{3/2}}
\ee

\begin{figure}[htbp]
\begin{center}
\includegraphics[height=6.5cm]{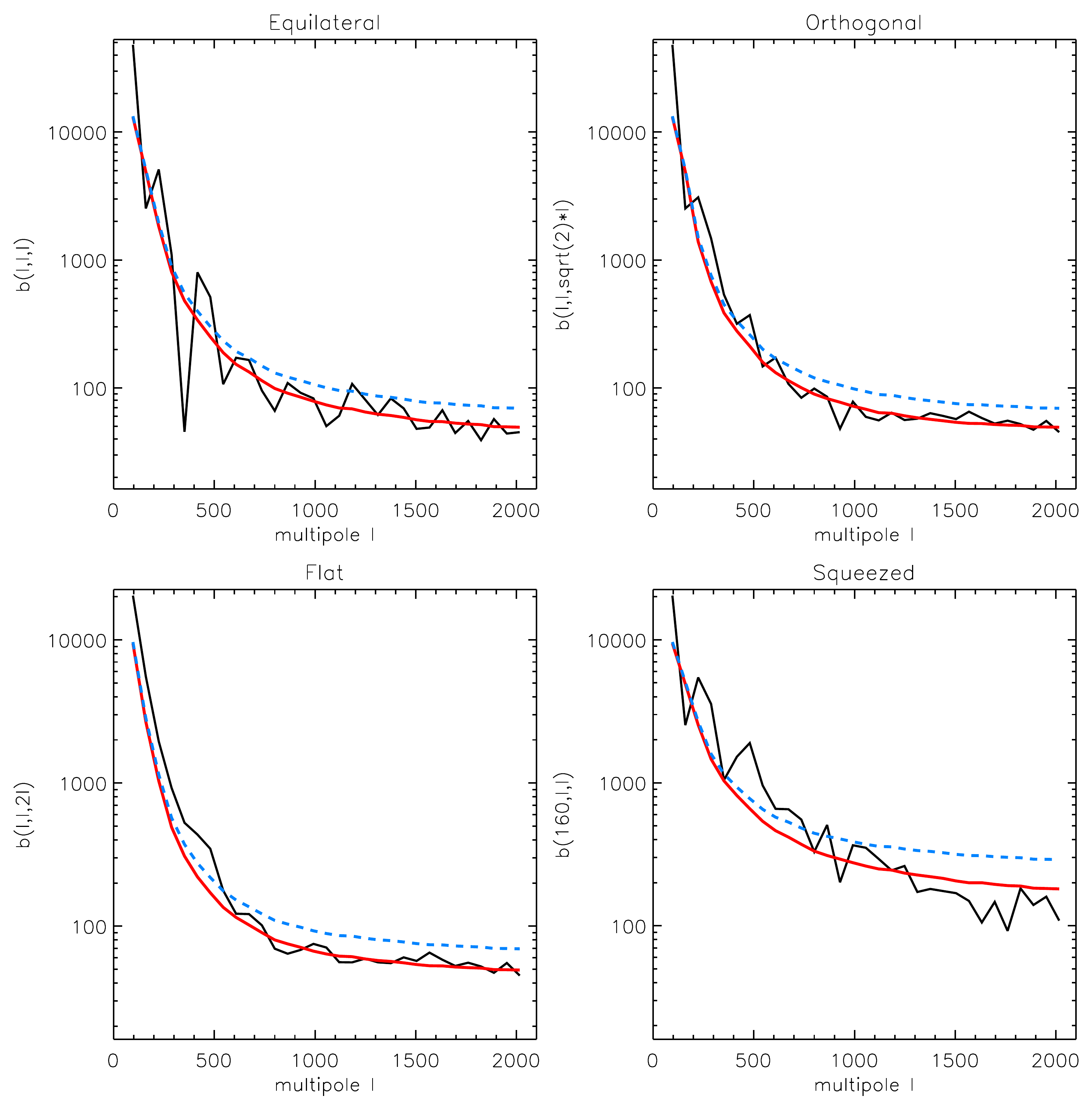} \hspace{2pc}
\begin{minipage}[b]{7cm}
\caption{\label{Fig:speconfirad350} IR+radio bispectrum at 350 GHz in some configurations. Black line is the measured bispectrum from the simulations, red line is the result of prescription, blue dashed line is the case where radio and IR populations are assumed correlated with each other.}
\end{minipage}
\end{center}
\end{figure}

We used publicly available simulations of these two populations between 30 and 350 GHz by \cite{Sehgal2010} to compare the measured bispectrum with our prescription. Fig.\ref{Fig:speconfirad350} shows the measured and predicted bispectra at 350 GHz in the case of superposition of IR and radio sources.

\vspace{-0.3cm}

We found a good agreement, within cosmic variance error bars, between the prescription and the measured bispectrum at all frequencies and in all cases: radio sources alone, IR alone, radio and IR together. It is also noticeable that the IR clustering increases NG by several order of magnitudes at large angular scales -- as compared to unclustered case. While the radio bispectrum is flat, the IR one exhibits a configuration dependence due to clustering (see at 148 GHz visualisation in Fig.\ref{Fig:triparametparamIR}, right panel). The IR bispectrum peaks in the squeezed triangles --as predicted by our prescription-- with  a slight flattening at high multipoles due to shot-noise, especially at the higher frequencies.

\section{Estimating the contamination to primordial NG}
We used the KSW estimator described previously (Eq.\ref{Eq:KSW}, \cite{Komatsu2005}) on the simulated maps, to estimate the bias of the primordial local NG parameter $\fnl$ due to foreground signals. We computed this bias for two angular resolution (WMAP-like, Planck-like), with and without masking sources above the Planck ERCSC catalog \cite{Planck-Collaboration-ERCSC} flux cuts. Results are presented in Table \ref{dfnlradtable}\&\ref{dfnlirtable} for radio and IR sources respectively.

\begin{table}
\centering
\caption{\label{dfnlradtable} Bias from radio sources to local NG parameter $\fnl$, depending on frequency, angular resolution, and masking or not detectable sources}
\begin{tabular}{@{}l @{}@{}c@{}c@{}c@{}c@{}c@{}c@{}}
\br
\multicolumn{7}{c}{without flux cut}\\
\mr
 $\nu$ (GHz) & $\ $30$\ $ & 90$\ $ & 148$\ $ & 219$\ $  & 277$\ $ & 350\\
\hline
$\ell_\mathrm{max}=50$ & -4.2& -0.0025 & -0.00037 & -0.00021 & -0.00027 & -0.00068\\
$\ell_\mathrm{max}=700$ & 3850& 2.5 & 0.38 & 0.21 & 0.27 & 0.65\\
$\ell_\mathrm{max}=2048\ $ & 177000& 117& 18& 9.7 & 12& 30 \\
\mr
\multicolumn{7}{c}{with flux cut}\\
\mr
$\ell_\mathrm{max}=700$ & 108 &Ê0.17Ê& 0.0071Ê& 0.0031Ê& 0.0035Ê& 0.0064\\
$\ell_\mathrm{max}=2048\ $ & 4930Ê& 7.5Ê& 0.31Ê& 0.14 & 0.16Ê& 0.29 \\
\br
\end{tabular}

\vspace{0.1cm}

\centering
\caption{\label{dfnlirtable} Bias from IR sources to local NG parameter $\fnl$}
\begin{tabular}{@{}l@{}@{}c@{}c@{}c@{}c@{}c@{}c@{}}
\br
\multicolumn{7}{c}{without flux cut}\\
\mr
 $\nu$ (GHz) & 30 & 90 & 148 & 219$\ $  & 277$\ $ & 350\\
\hline
$\ell_\mathrm{max}=50$ & $\,$-$3.5 \!\cdot\! 10^{-8}$ & $\,$-$5.9 \!\cdot\! 10^{-6}$ & $\,$-$9.2 \!\cdot\! 10^{-5}$ & -0.0027 & -0.023 & -1.0\\
$\ell_\mathrm{max}=700$ & $\,$-$1.3 \!\cdot\! 10^{-6}$ & -0.00019 & -0.0033 & -0.063 & -0.55 & -9.0 \\
$\ell_\mathrm{max}=2048\,$ & $\,$-$1.8 \!\cdot\! 10^{-5}$& -0.0026 & -0.039 & -0.68 & -4.8 & -67\\
\mr
\multicolumn{7}{c}{with flux cut}\\
\mr
$\ell_\mathrm{max}=700$ & $\,$-$1.3 \!\cdot\! 10^{-6}$ & -0.00019 & -0.0033 & -0.078 & -0.74 & -11\\
$\ell_\mathrm{max}=2048\,$ &  $\,$-$1.8 \!\cdot\! 10^{-5}$& -0.0026 & -0.039 & -0.67 & -6.3 & -66 \\
\br
\end{tabular}
\end{table}

The relative error bars of these biases are of 1.5-2.5\% for radio sources and 3 to 7\% for IR sources.\\
We find that for both populations the bias increases at higher maximum multipoles, as the CMB signal plummets due to Silk damping. The radio bias is positive and maximal at low frequencies, while the IR bias is negative and peaks at the higher frequencies. At a WMAP-like resolution, except at 30 GHz, these biases can be neglected compared tho the expected error bars on $\fnl$ due to cosmic variance. However at Planck-like resolution, these biases become non-negligible ; while the radio bias is very efficiently reduced by masking detectable sources, it is not the case for IR sources where the bias is unaffected. Indeed the IR population, and especially the clustered galaxies, is dominated by faint sources much below the detection limit. 

\section*{References}
\bibliographystyle{iopart-num}
\bibliography{proceeding_fabien-lacasa-wtemplate}

\end{document}